\documentclass[conference]{IEEEtran}
\usepackage{blindtext, graphicx}
\ifCLASSINFOpdf
\else
\fi
\usepackage[cmex10]{amsmath}
\begin{document}
%
\title{Inference of principal components of noisy correlation matrices with prior
information}

\author{\IEEEauthorblockN{R\'emi Monasson}
\IEEEauthorblockA{Laboratory of Theoretical Physics, CNRS \& Ecole Normale Sup\'erieure,
PSL Research, 24 rue Lhomond, Paris, France \\
Email: see http://www.physique.ens.fr/$\sim$monasson}
}


%


\maketitle

\begin{abstract}
The problem of infering the top component of a noisy sample covariance matrix with prior information about the distribution of its entries is considered,  in the framework of the spiked covariance model. Using the replica method of statistical physics the computation of the overlap between the top components of the sample and population covariance matrices is formulated as an explicit optimization problem for any kind of entry-wise prior information. The approach is illustrated on the case of top components including large entries, and the corresponding phase diagram is shown. The calculation predicts that the maximal sampling noise level at which the recovery of the top population component remains possible is higher than its counterpart in the spiked covariance model with no prior information.
\end{abstract}


%
\IEEEpeerreviewmaketitle

\section{Introduction}
In the era of big data inferring features of complex systems, characterized by many degrees of freedom, is of crucial importance. The high-dimensional setting, where the number of features to extract is not small compared to the number of available data, makes this task statistically or computationally hard. One case of practical interest is the inference of the largest component (eigenvector) of correlation matrices. Consider $T$ independently drawn observations of $N$ interacting Gaussian variables, {\em i.e.} such that the population covariance matrix $C$ is not the identity matrix. If $T$ is much larger than $N$ the empirical covariance matrix $\hat  C$ computed from the $T$ observations converges to $C$, and recovering the top eigenvector is easy. The case where both $N,T$ are large (sent to infinity at fixed ratio $r=N/T$) has received a lot of attention, both theoretically and practically \cite{hdi}. From a theoretical point of view, it has been shown, in the case of a covariance matrix $C$ with one (or few compared to $N$) eigenvalues larger than unity, say, $\gamma$, that recovery is possible if $r$ is smaller than the critical value $r_c=(\gamma-1)^2$ \cite{hoyle,bbp}. For larger sampling noise ($r>r_c$), the top eigenvector of $\hat C$ is essentially orthogonal to the top component of $C$, and is therefore not informative. It is reasonable to expect that the situation will improve in the presence of additional, prior information about the structure of the top component to be recovered, and that recovery will be possible even when $r$ is (not too much) larger than $r_c$. That this is indeed the case has been rigorously shown when all entries are nonnegative \cite{montanari}, and is supported by strong numerical evidence when the top component is known to have large entries (finite as $N\to\infty$) \cite{villamaina}. In the present work, using techniques from statistical physics we propose explicit conjectures about the critical noise level and its dependence on the signal eigenvalue ($\gamma$) and on prior knowledge. The framework is general and can be applied to any kind of entry-wise prior probability, {\em i.e.}  factorized over the entries $\xi_i$ of the top component $\boldsymbol\xi$. We show how rigorous results in the nonnegative case of \cite{montanari} are recovered, and present new results for the  large entry prior. 

The motivation to consider the latter prior stems from computational biology, more precisely, from the study of coevolution between amino acids in protein families. Sequences of proteins diverged from a common ancestor widely differ across many organisms, while the protein structure and function are often very well conserved. The constraints induced by structural and functional conservation manifest themselves as correlations between amino acids (the $N$ variables, where $N$ is the protein length) across the different organisms (the $T$ observations). Recently, it was shown that the eigenmodes $\boldsymbol\xi$ of the amino-acid covariance matrix corresponding to {\em low} eigenvalues were informative about three-dimensional contacts on the protein structure \cite{cocco13}. These modes show large entries on the protein sites and amino-acid types in contact; as the other entries contain diffuse, non-structural signal \cite{lp}, the components $\boldsymbol\xi$ cannot be thought of as being sparse. The presence of large entries in structurally-informative components was empirically assessed through the so-called inverse participation ratio, $\sum _i \xi_i^4$ (for normalized $\boldsymbol\xi$), a quantity that remains finite for components with (few) large entries and otherwise vanishes for $N\to\infty$. We hereafter use this quantity as a prior over the components to facilitate their recovery.

\section{Probabilistic framework}

\subsection{Spiked covariance model}
We consider the popular Spiked Covariance Model, in which the entries of $N$-dimensional  vectors, ${\bf x}=(x_1,x_2,...,x_N)$, are Gaussian random variables with zero means and population covariance matrix $\bf C$. All eigenvalues of $\bf C$ but one are equal to unity, while the remaining eigenvalue is $\gamma\ne1$, with associated eigenvector $\bf u$. As usual we choose $\gamma>1$ but our results could be transposed to the case $\gamma<1$ with minor modifications. We draw $T$ independent samples ${\bf x}^{t}, t=1,2,...,T$, and define the sample covariance matrix $\hat{\bf C}$, with entries $\hat C_{ij}=\frac 1T \sum _t x_i^t x_j^t$. The top eigenvector of $\hat{\bf C}$ is denoted by $\boldsymbol\xi$. In the double limit $N,T\to \infty$ at fixed ratio $r=N/T$, there exists a phase transition at a critical value of the sampling noise $r_c = (\gamma-1)^2$ separating the {\em high-noise regime}, $r>r_c$, in which $\boldsymbol\xi$ and $\bf u$ have asymptotically zero squared dot product, and the {\em low-noise regime}, $r<r_c$, where the squared dot product between $\boldsymbol\xi$ and $\bf u$ is strictly positive with high probability \cite{hoyle,bbp}. 

\subsection{Likelihood of principal component $\boldsymbol \xi$}
The sample covariance matrix $\hat {\bf C}$ obeys a Wishart distribution, determined by $\bf C$, $N$ and $T$. Using Bayes formula we may write the likelihood (density of probability) for the normalized top component $\boldsymbol\xi$ of $\hat{\bf C}$ as follows
\begin{equation}
\rho (\boldsymbol\xi) \propto \exp\left( \frac{r\, \beta}2\, \sum_{i,j} \xi_i \, \hat C_{ij}\, \xi_j\right) \delta (\boldsymbol\xi^2-1)\ ,
\end{equation}
up to a normalization coefficient. Parameter $\beta$ above is equal to $\beta_{Bayes}=1-\frac 1\gamma$. However, it is convenient to consider $\beta$ as a free parameter. The $\beta\to\infty$ limit corresponds to Maximum Likelihood inference, while working at low values of $\beta$ may be useful to ensure rapid mixing of Monte Carlo Markov Chain sampling of distribution $\rho$, especially in the presence of prior information, see below. 

\subsection{Prior information over $\boldsymbol \xi$}
\begin{figure}[!t]
\centering
\includegraphics[width=3.7in]{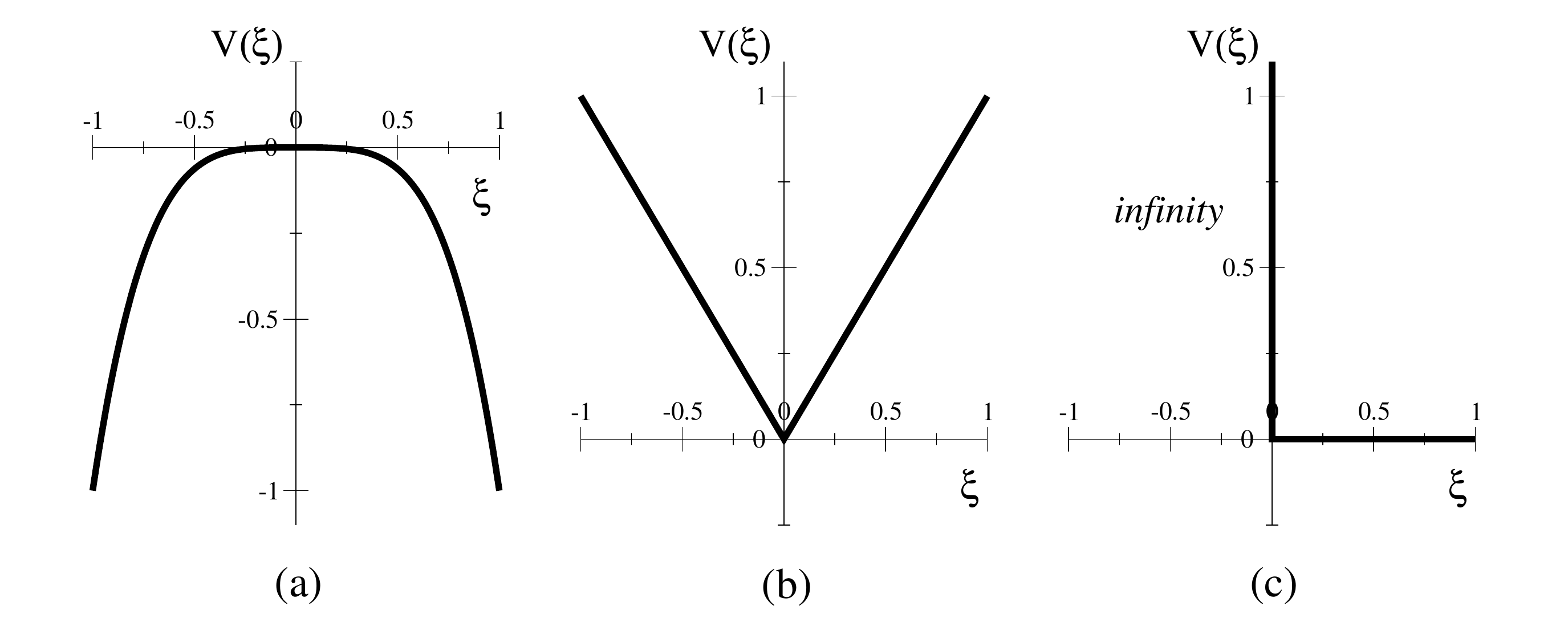}
\caption{Three potentials $V(\xi)$ corresponding to three prior information about the entries $\xi$ of the principal component: (a) large entries are present, (b) $L_1$ penalty favoring zero entries, (c) all entries are nonnegative. }
\label{fig_pot}
\end{figure}

We now assume that prior information over the population eigenvector $\bf u$ is available under the form of a potential $V$ acting on the entries of $\boldsymbol\xi$. The posterior distribution over the top component now reads
\begin{equation}\label{rho2}
\rho (\boldsymbol\xi) \propto \exp\left( \frac{r\, \beta}2\, \sum_{i,j} \xi_i \, \hat C_{ij}\, \xi_j -\sum_i V(\xi_i)\right) \delta (\boldsymbol\xi^2-1)\ ,
\end{equation}
up to a normalization coefficient. Three choices for the potential $V$ are shown in Fig.~\ref{fig_pot}. Motivated by previous works on protein sequence analysis, see Introduction, we will hereafter mostly concentrate on $V(\xi)=-V_0\; \xi^4$, with $V_0\ge 0$ (Fig.~\ref{fig_pot}(a)). This potential favors the presence of large entries in the top component,  but does not rule out the existence of many entries with small magnitude (typically, of the order of $N^{-1/2}$). It is therefore different from sparsity-enforcing potentials, such as $V(\xi)\propto |\xi|$ in Fig.~\ref{fig_pot}(b). Exact results for the location of the transition in the nonnegative case (Fig.~\ref{fig_pot}(c)) were recently derived \cite{montanari}. Our formalism finds back those results, and can be applied to any potential $V$ as shown below.

\section{Calculation of phase diagram}

\subsection{General replica calculation}
We assume that the logarithm, divided by $N$, of the normalization coefficient of $\rho$ in Eq.~(\ref{rho2}),
\begin{equation}
Z(\hat{\bf C}) = \int_{ \boldsymbol\xi^2=1} d\boldsymbol\xi\; \exp\left( \frac{r\, \beta}2\, \sum_{i,j} \xi_i \, \hat C_{ij}\, \xi_j -\sum_i V(\xi_i)\right) 
\end{equation}
is concentrated around its expectation value $L\equiv \frac 1N E_{\hat {\bf C}}[\log Z]$ in the $N,T\to\infty$ limit, and compute the latter with the help of the replica method, a non rigorous technique commonly used in the statistical physics of disordered systems \cite{replicas}, see \cite{ganguli,obuchi,krzakala} for recent applications to related high-dimensional inference problem. We obtain
\begin{equation}\label{logz}
\begin{split}
&L(r,\gamma,\beta) = \text{Extr} \bigg\{ -\frac 1{2r}  \log\bigg(\frac 1\beta-q_0+q_1\bigg)  -q_0\hat q_0 \\
&+ \frac{q_1}{2r(\frac 1\beta -q_0+q_1)} + \big(\hat q_0-\hat q_1 \big)\big(1+(\gamma-1) p^2\big) \\
& + \hat \mu + q_1\hat q_1 +p\hat p  -\frac 12 \log \hat \mu + \frac {\hat q_1}{2\hat\mu} +\tilde L(\hat \mu,\hat p, {\bf u}) \bigg\} \ ,
\end{split}
\end{equation}
where the Extremum is computed over the order parameters  $p^2= \frac 1N E_{\hat{\bf C}} [\langle ( \sum_{i} u_i  \xi_i)^2\rangle_{\hat{\bf C}}]$, $q_0=\frac 1N \sum_{i,j}C_{ij} E_{\hat {\bf C}}[\langle \xi_i \xi_j\rangle_{\hat{\bf C}}]$, $q_1=\frac 1N\sum_{i,j}  C_{ij} E_{\hat{\bf C}} [\langle \xi_i \rangle_{\hat{\bf C}}\langle \xi_j\rangle_{\hat{\bf C}}]$, and the conjugated Lagrange multipliers $\hat q_0,\hat q_1,\hat \mu$. Here, $\langle\cdot\rangle_{\hat{\bf C}}$ denotes the expectation over the $\rho$ distribution over $\boldsymbol\xi$ in Eq.~(\ref{rho2}). The term $\tilde L$ in Eq.~(\ref{logz}) depends on the prior potential $V$ and on the structure of the normalized population top component ${\bf u}$, more precisely, on how its entries scale with $N$.  We now consider two cases of interest.

\subsection{Case of nonnegative entries}
We assume first that the components of $\bf u$ scales as $\frac{\tilde u_i}{\sqrt N}$, with $\tilde u_i$ finite, and denote by $\varphi(\tilde u)$ the distribution of the $\tilde u_i$. We focus on the nonnegative entry prior, for which $V(\xi)=+\infty$ for $\xi<0$ and 0 for $\xi\ge0$. We obtain
\begin{equation}\label{logz2}
\begin{split}
&\tilde L= \frac{\hat p^2}{2\hat\mu} + \frac 1{2\hat \mu} \int d\tilde u\, \varphi(\tilde u)\bigg[\log \text{erfc}\left(\frac{-\hat p\, \tilde u -z \sqrt{2\hat q_1}}{2\sqrt{\hat \mu}}\right)\bigg]_z \ ,
\end{split}
\end{equation}
where erfc is the complementary error function, and $[F(z)]_z$ denotes the average of $F(z)$ over the Gaussian measure, $e^{-z^2/2}/\sqrt{2\pi}$. After some elementary algebra, we obtain the expression for the overlap 
\begin{equation}
p = \frac{ [\tilde u (x\,\tilde u+z)]_z^+}{[(x\,\tilde u+z)^2]_z^+}\ \text{with}\ 
x=\frac{\hat p}{\sqrt{2\hat q_1}}=\frac{(\gamma-1)\, p}{\sqrt{r(1+(\gamma-1)p^2)}}
\end{equation}
where  $[F(z))]_z^+=[\max(F(z),0)]_z$. These equations correspond to Eqs.~(7), (8), (9), (21) and (23) of \cite{montanari}.

\subsection{Case of large entries}\label{poiuy}

We now assume that ${\bf u}$ has only $K$ `large' entries, $u_1,...,u_K$, different from zero in the $N\to\infty$ limit (with finite $K$), and that the other entries decay fast enough with $N$, {\em e.g.} are of the order of $\frac 1{\sqrt N}$.  Then
\begin{equation}\label{logz3}
\begin{split}
&\tilde L=   -\sum_{i=1}^K \min_{-1\le\xi\le 1} \big\{ \hat \mu \,\xi^2 +\hat p\, u_i\, \xi +V(\xi)\big\}\ ,
\end{split}
\end{equation}

While the above formula is valid for generic $V$ we restrict ourselves, in the remaining part of this article, to the potential $V(\xi)=-V_0\; \xi^4$. Furthermore, we assume  that the $K$ finite entries of $\bf u$ are all equal to  $u=\frac 1{\sqrt K}$; The calculation can be easily extended to $u<\frac 1{\sqrt K}$, or to the case of nonhomogeneous entries $u_i$. In addition we assume that (A1) all $\xi_i$s take identical values in Eq.~(\ref{logz}) (homogeneous regime); (A2) $\boldsymbol\xi$ has no large (finite when $N\to\infty$) entry $\xi_i$ on sites such that $u_i=0$. The validity of these assumptions will be discussed in the next Section.

After some elementary algebra the extremization conditions reduce to the following set of $K+1$ coupled equations over $\hat \mu$ and the first $K$ entries of $\boldsymbol\xi$:
\begin{equation} \label{ppo}
\begin{split}
& \xi_i -\frac{2\,V_0}{\hat\mu}\, \xi_i^3=\frac{(\gamma-1) p}{r\sqrt K\left(\frac{2\hat \mu}{\beta}-1\right)}\, \quad (i=1,...,K)\ ,\\
& \sum_{i=1}^K \xi_i^2 = 1-\frac{1}{2\hat \mu} - \frac{1+(\gamma-1) p^2 - \frac 1{2\hat\mu}}{r \left(\frac{2\hat \mu}{\beta}-1\right)^2}
\end{split}
\end{equation}
where 
\begin{equation}
p=\frac 1{\sqrt K} \sum_{i=1}^K \xi_i \ .
\end{equation}
Note that the $K$ variables $\xi_i$s obey the same cubic equation and, hence, can take at most three different values as $i$ varies. 

The equations above admit the solution $p=\xi_i=0,\hat\mu=\frac 12$, corresponding to the `unaligned' phase. In addition, in some well-defined regions of the four-dimensional control parameter space $(r,\gamma, \beta,V_0)$ solutions with $p\ne0$ may be found. We give in Section IV below results for three cases: (A) no prior ($V_0=0$); (B) weak exploitation of many data with prior information (small $\beta, r$ for finite $V_0$); (C) maximum {\em a posteriori} inference (finite $r$, and infinite $\beta$ and $V_0$ at fixed ratio $V_0/\beta$).

\section{Results for `large entry' prior}

\subsection{Warm-up: no prior}
We start with the case $V_0=0$.  Extremization conditions over the parameters in Eq.~(\ref{logz}) give the value of the squared overlap $p^2$ between ${\bf u}$ and $\boldsymbol\xi$ for any $\beta$. We find: $p^2=0$ for $r> r_c=(1-\gamma)^2$ whatever the value of $\beta$, and, for $r<r_c$, 
\begin{equation}
p^2=\left(1-\frac{\beta(r)}{\beta}\right)\left( \frac{1-\frac{r}{r_c^2}}{1+\frac r{r_c}}\right)\text{if}\ \beta >\beta(r)\equiv \frac{r}{r+\gamma-1}.
\end{equation}
Those expressions perfectly agree, in the $\beta\to\infty$ limit,  with known results for the spiked covariance model  \cite{hoyle,bbp}. In addition our formalism gives access to the value of $p^2$ for finite $\beta$. Note that, for $r<r_c$, inference of the direction of ${\bf u}$ is possible, {\em i.e.} $p^2>0$, even for $\beta \le  \beta_{Bayes}$ (but larger than $\beta(r)$). At the critical noise, $\beta(r_c)=\beta_{Bayes}$.

\subsection{Inference at low $\beta$ with prior information}
The above results imply that, in the absence of prior information ($V_0=0$), inference of the top component direction is possible at low $\beta\to 0$ provided the sampling noise $r$ is smaller than $\beta /(1-\gamma)$. In other words, when both $\beta$ and $r$ tend to zero with a fixed ratio $\tilde\beta=\beta/r$, the aligned and not-aligned phases correspond, respectively, to $\tilde \beta > \tilde\beta_c (0) = \frac 1{\gamma-1}$, and $\tilde \beta < \tilde \beta_c (0)$. 

\begin{figure}[!t]
\centering
\includegraphics[width=3.3in]{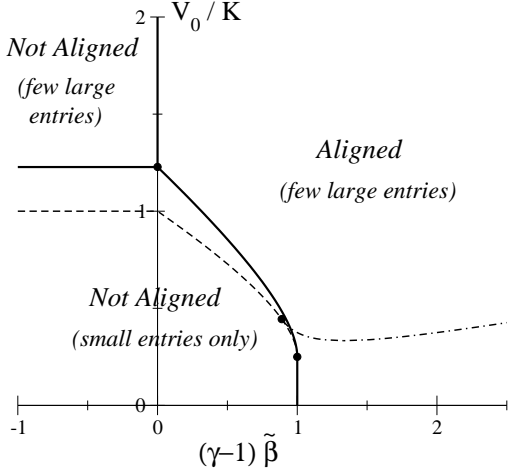}
\vskip -.3cm
\caption{Phase diagram of top component recovery in the ($\tilde \beta, V_0)$ plane; axis are rescaled by $(\gamma-1)$ and $\frac 1K$ factors, where $K$ is the number of nonzero components in ${\bf u}=\frac 1{\sqrt K}(1,1,...,1,0,0,...,0)$. Dots locate the points $(1,\frac 14)$, $(\frac 89,\frac 49)$ and $(0,1.228...)$, see text. Transitions between phases are shown by full lines. The dashed lines show the limit of existence of the Aligned phase, while the dot-dashed lines separate the regions with (above) and without (below line) homogeneity breaking among the $K$ large entries of $\boldsymbol\xi$.}
\label{fig_pd}
\end{figure}

\begin{figure}[!t]
\centering
\includegraphics[width=3.in]{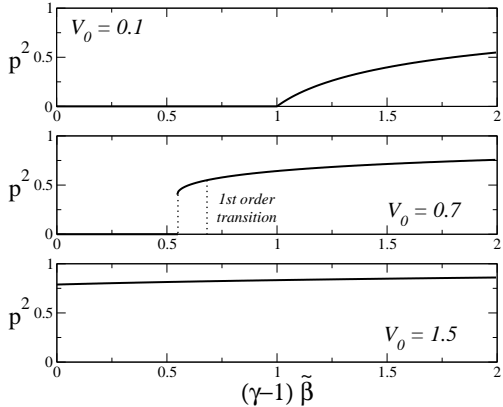}
\vskip -.5cm
\caption{Average squared overlap between top components of the population and sample covariance matrices, $p^2\equiv E_{\hat{\bf C}} [({\bf u}\cdot \boldsymbol\xi )^2]$, vs. control parameter $(\gamma-1)\tilde \beta$ for three prior strengths and $K=1$. }
\label{fig_p2}
\end{figure}

We expect the critical ratio $\tilde \beta_c(V_0)$ to be a drecreasing function of the prior strength $V_0$, as stronger prior should facilitate the recovery of the large-entry top component $\bf u$. Resolution of Eq.~(\ref{ppo}) gives the phase diagram shown in Fig.~\ref{fig_pd}. Several regimes can be identified, depending on $V_0$:
\begin{itemize}
\item For $0<V_0\le \frac K4$, the critical ratio $\tilde \beta_c(V_0)$ remains unchanged, see Fig.~\ref{fig_pd}, and equal to $\frac 1{\gamma-1}$. As $\tilde \beta$ crosses this critical value the overlap $p$ continuously increases from 0 to a positive value, see Fig.~\ref{fig_p2}.
\item For $\frac K4 < V_0\le K$, the aligned phase ($p\ne 0$) exist for $\tilde \beta  >  \tilde \beta_{c,1} = \frac{4 \sqrt{V_0/K}\big(1-\sqrt{V_0/K}\big)}{\gamma-1}$, see dashed line in Fig.~\ref{fig_pd}. As $\tilde \beta$ crosses $\tilde \beta_{c,1} $ the squared overlap $p^2$ discontinuously jumps from 0 to $1-\frac 1{2\sqrt{V_0}}>0$, see Fig.~\ref{fig_p2}.
\item For $\frac K4 < V_0\le V_0^+\times K$, the aligned phase ($p\ne 0$) is thermodynamically stable, meaning that the value $L_+$ of $L$ in Eq.~(\ref{logz}) attached to this phase is larger than the one of the nonligned phase ($p=0$), $L_0=\frac 12 (1+\tilde \beta)$, for $\tilde \beta  >  \tilde \beta_{c,2} $, see full line in Fig.~\ref{fig_pd}. The value of $\tilde \beta_{c,2} $ and of the overlap $p_2$ at the phase transition are the roots of the two coupled equations
\begin{equation} \label{ppw}
\begin{split}
& 0= \frac{p_2^2}{1-p_2^2} + \frac 12 \log\big(1-p_2^2\big) -\frac{V_0}{K} \, p_2^4 \ , \\
& \tilde \beta_{c,2} = \frac 1{1-p_2^2} - 4 \frac{V_0}K\,p_2^2 \ ,
\end{split}
\end{equation}
where the first equation actually implements the condition $L_+=L_0$. The phase transition is illustrated in the middle panel of Fig.~\ref{fig_p2} for a specific value of $V_0$. The value of $V_0^+\simeq 1.227703...$ is defined from
\begin{equation}
V_0^+ = \min \big\{ V_0 \ \text{s.t.} \max_{\xi  \ne 0} \big( \log(1-\xi^2) +2V_0\, \xi^4\big) > 0\big\}
\end{equation}
\item For $V_0\ge V_0^+\times K$ the prior strength is so strong that the inferred component $\boldsymbol\xi$ has few large entries whatever the value of $\beta$. For $\beta>0$ it is aligned ($p\ne 0$) with $\bf u$ (Fig.~\ref{fig_p2}), while for $\beta<0$, it is not, see Fig.~\ref{fig_pd}.
\end{itemize}

Assumption (A1), see Section \ref{poiuy},  is trivially valid for $K=1$, but is not necessarily correct for $K\ge 2$ and strong prior strength, for which we expect that $\boldsymbol\xi$ will condensate and one component, say, $\xi_1$, will be larger than the other components, say, $\xi_i$, with $i=2,...,K$ (nonhomogeneous regime). The transition line between these two regimes is identified upon imposing that the cubic equation over $\xi_i$ in Eq.~(\ref{logz}) admits a two-fold degenerate solution $\xi$, that is, $6V_0\,\xi^2=\hat\mu$. The transition line is plotted in the phase diagram of Fig.~\ref{fig_pd} (dot-dashed line), and ends up in the point of coordinate $(\frac 49,\frac 89)$ lying on the existence line (dashed line). 

\subsection{Inference at high $\beta$  with strong prior information}
\begin{figure}[!t]
\centering
\includegraphics[width=2.8in]{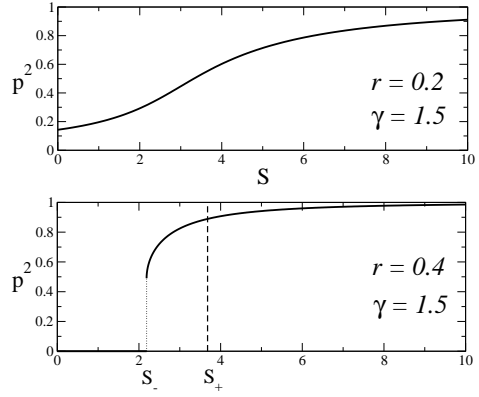}
\vskip -.3cm
\caption{Average squared overlap between top components of the population and sample covariance matrices, $p^2$,  vs. slope $S$ for  sampling noises $r=0.2 $ (top) and 0.4 (bottom). Note the presence of the discontinuous transition at $S_-\simeq 2.19$ in the latter case. The randomly condensed solution appears for $S>S_+\simeq 3.68$.  Here, $r_c=(\gamma-1)^2=0.25$.}
\label{fig_S1}
\end{figure}

\begin{figure}[!t]
\centering
\includegraphics[width=3.7in]{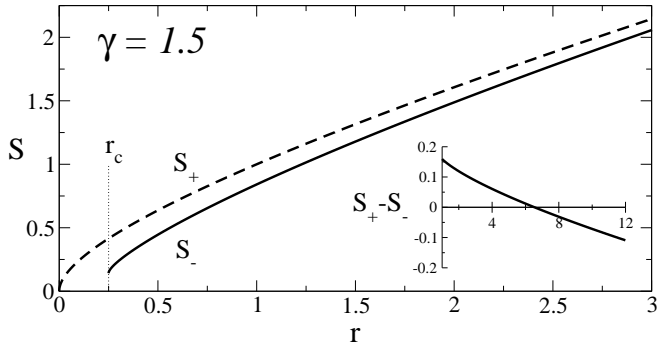}
\vskip -2.6cm
\caption{Behaviours of $S_-$ (full line) and $S_+$ (dashed line) as functions of the sampling noise $r$. Insert: $S_+-S_-$ vs. $r$. The difference vanishes in $r_d\simeq 6.54$. Here, $r_c=0.25$, {\em i.e.} $\gamma=1.5$.}
\label{fig_S2}
\end{figure}

We now focus on MAP inference at finite sampling noise $r$, whereas $\beta$ and $V_0$ are both sent to infinity, with a fixed ratio $S=r\,V_0/(\beta)$. Parameter $S$, hereafter referred to as the slope, controls the relative magnitude of the $\hat C$-dependent and prior terms in $\rho$, see Eq.~(\ref{rho2}), while the multiplicative factor $r$ is introduced in the definition of $S$ to compensate for the explicit dependence upon $r$ in $\rho$. For simplicity we present results for $K=1$ only, the extension to larger $K$ being rather straightforward. Equations~(\ref{ppo}) admit the solution $p=0,\hat \mu=\frac 12$, and another solution, with $\hat\mu\to\infty $ as $\beta,V_0\to\infty$, with ratios $\beta/\hat\mu$, $V_0/\hat\mu$ having finite limits. After some simple algebra we obtain the following expresson for the slope as a function of the squared overlap for the latter solution:
\begin{equation}
S(p^2)= \frac{\big(r-(\gamma-1) y\big)\big( 1+y\big)}{4\,p^2\, y}\ \text{with}\ y =\sqrt{\frac{r(1-p^2)}{1+(\gamma-1)p^2}}.
\end{equation}
We show in Fig.~\ref{fig_S1} the representative curve of $p^2$ vs. the slope $S$, for $r$ below and above the critical noise level, $r_c=(\gamma-1)^2$, in the absence of prior. For $r<r_c$ the squared overlap is an increasing function of $S$, starting from a non zero value for $S=0$: the population eigenvector direction can be estimated without prior at low sampling noise \cite{bbp}, but the overlap is increased in the presence of prior. For $r > r_c$ a discontinuous jump is observed from $p=0$ to $p>0$ at a critical value of the slope,
\begin{equation}
S_- = \min_{p^2>0} S(p^2) \qquad (r>r_c) \ ,
\end{equation}
while the overlap further increases as $S$ exceeds $S_-$. Remarkably, even for large sampling noise values, the presence of a sufficiently strong prior allows us to infer ${\bf u}$. The value of $S_-$ as a function of the noise level $r$ is shown in Fig.~\ref{fig_S2}; for large noise levels we have $S_-=\frac r4+\frac 3{4^{4/3}} \,\gamma^{1/3}\, r^{2/3}+ O(r^{1/3})$.

This aligned phase competes with a nonaligned, but condensed phase, in which assumption (A2), see Section \ref{poiuy}, is violated. In other words, for $S$ and $r$ sufficiently large, $\boldsymbol\xi$ has few large entries ($\xi_j^2>0$ in the $N\to\infty$ limit), but not along the directions $i$ corresponding to the large components of $\bf u$; hence, $p^2=0$. To describe this new phase we set $u_i$ to 0 in the expression for $\tilde L$ in Eq.~(\ref{logz3}). The corresponding optimization equations  can be solved in the $\beta,V_0\to\infty $ limit, with the result that the nonaligned, condensed regime exists for $S$ larger than
\begin{equation}
S_+ = \min_{0<b<\frac{\sqrt{r}}{1+\sqrt{r}}} \bigg[ \frac {r^2(1-b)^2}{4b\big( r(1-b)^2-b^2\big)}\bigg] \ .
\end{equation}
The value of $S_+$ as a function of the noise level $r$ is shown in Fig.~\ref{fig_S2}. For small $r$ (slightly above $r_c$) we observe that $S_+$ is larger than $S_-$, as intuitively expected: it is favorable to condense $\boldsymbol\xi$ along the direction of $\bf u$ rather than any other direction. It can be checked that the value of $L$ in Eq.~(\ref{logz}) is higher for this phase than for the aligned condensed phase. Hence, as soon as $S$ exceeds $S_+$ the overlap $p$ vanishes.

\begin{figure}[!t]
\centering
\includegraphics[width=3.4in]{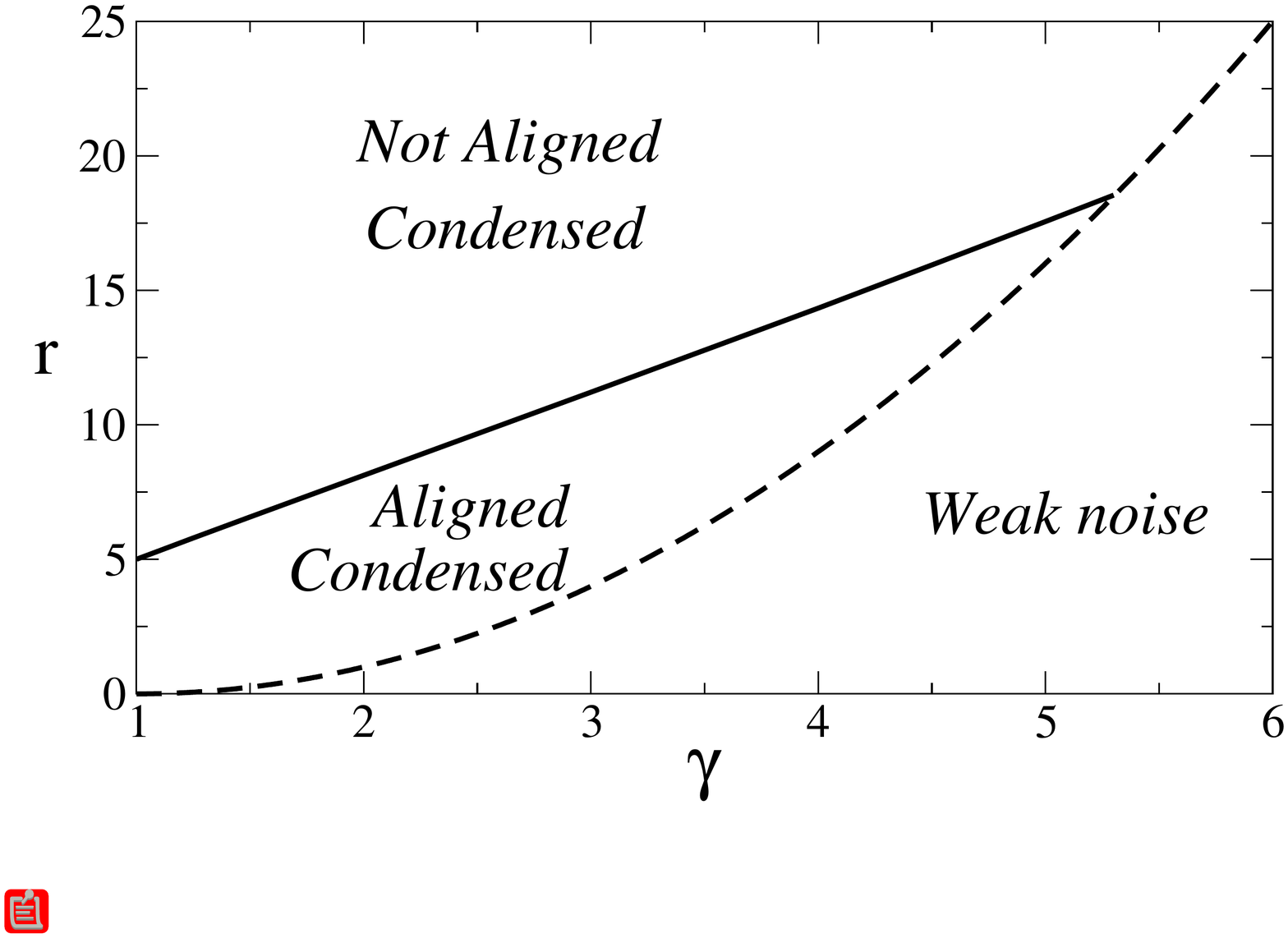}
\vskip -1.9cm
\caption{Replica-symmetric phase diagram of top component inference with large-entry prior in the $(\gamma, r)$ plane. The dashed line $r_c=(\gamma-1)^2$ divides the plane into the weak noise region (below line), where inference is possible withour prior, and the strong noise region (above line). The full line shows the value of $r_d$, at which $S_-=S_+$, as a function of $\gamma$. In between the dashed and full lines, inference of the top component is possible in the presence of a prior with appropriate strength. The two lines merge for $\gamma\simeq 5.3$, $r\simeq 18.5$. Here, $K=1$.}
\label{fig_S3}
\end{figure}

For large noise levels, however, we have $S_+=\frac r4+\frac 3{4^{4/3}} r^{2/3}+ O(r^{1/3})$, which is asymptotically smaller than $S_-$, see insert of Fig.~\ref{fig_S2}. Indeed, the threshold slopes $S_-$ and $S_+$ cross at a well-defined value of the noise, $r_d$, which depends on the top eigenvalue $\gamma$. We show in Fig.~\ref{fig_S3} the behaviour of $r_d$ vs. $\gamma$, and compare it to the critical noise $r_c$ in the absence of prior. We observe the presence of a region in the $(\gamma,r)$ plane, above the critical line $r_c$, where the direction of $\bf u$ can be inferred thanks to the large-entry prior. Our replica symmetric theory predicts that the benefit of the prior does not extend to very large values of the signal eigenvalue $\gamma$ and of the noise $r$, see Fig.~\ref{fig_S3}.

\section{Conclusion}
The non rigorous calculations done in this paper suggest that inference of the top component of the population covariance matrix is possible in the presence of prior information, even above the critical noise level of the spiked covariance model, in agreement with rigorous results for the nonnegative case \cite{montanari} and numerical investigations for the large-entry case \cite{villamaina}. Many interesting questions have not been investigated in the present paper: how hard is the recovery problem from a computational point of view? Are there `dynamical' phase transitions separating subregions in the aligned phase, where the top component can be recovered in polynomial time or not? If so how do these line compare to the `static' critical lines derived in this paper? In addition it would be interesting to investigate the validity of the replica-symmetric hypothesis used to derive the results above \cite{replicas}. Though replica symmetry is generally expected to be correct for convex optimization problems what happens in nonconvex situations is not clear. For instance, inference of the top component with the nonnegative prior gives rise to a nonconvex optimization problem \cite{montanari}, but all rigorous results are exactly found back within our replica symmetric approach, see Section III.B. From this perspective it would be useful to investigate whether the present results are robust against replica symmetry breaking.


\section*{Acknowledgment}
I am grateful to M.R. McKay for the invitation to the Asilomar 2016 conference, and to S. Cocco and D. Villamaina for useful discussions. This work has benefited from the financial support of the CNRS Inphyniti Inferneuro project.

\ifCLASSOPTIONcaptionsoff
  \newpage
\fi

\end{document}